\documentstyle[12pt]{article}

\def\noi{\noindent}

\def\nqq{\hspace{-2em}}

\def\barr{\left(\begin{array}}
\def\earr{\end{array}\right)}
\def\beq#1{\begin{equation}\label{#1}}
\def\eeq{\end{equation}}
\def\ber#1{\begin{eqnarray}\label{#1} \nqq}
\def\eer{\end{eqnarray}}
\def\eern{\nonumber \end{eqnarray}}
\def\nn{\nonumber}
\def\mm{\\ \nqq}
\newcommand{\bear}[1]{\begin{eqnarray}\label{#1}}
\newcommand{\ear}{\end{eqnarray}}

\catcode`\@=11 \@addtoreset{equation}{section}\catcode`\@=12

\newcommand{\R}{\mbox{\bf R}}

\newcommand{\N}{\mbox{\bf N}}

\newcommand{\sign}{\mathop{\rm sign}\nolimits}

\newcommand{\sh}{\mathop{\rm sh}\nolimits}
\newcommand{\ch}{\mathop{\rm ch}\nolimits}

\newcommand{\eps}{\varepsilon}
\newcommand{\tri}{\triangle}

\newcommand{\p}{\partial}

\newcommand{\fnm}{\footnotemark}
\newcommand{\fnt}{\footnotetext}

\begin{document}

\begin{center}
\large\bf
 TODA P-BRANE BLACK HOLES AND POLYNOMIALS RELATED TO LIE ALGEBRAS
 \\[15pt]

\normalsize\bf V. D. Ivashchuk\fnm[1]\fnt[1]{e-mail:
  ivashchuk@mail.ru} and V. N. Melnikov\fnm[2]\fnt[2]{e-mail:
  melnikov@phys.msu.ru},

\vspace{0.3truecm}

 \it Center for Gravitation and Fundamental Metrology,
 VNIIMS, 46 Ozyornaya ul., Moscow 119361, Russia

 \it Institute of Gravitation and Cosmology,
 Peoples' Friendship University of Russia,
 6 Miklukho-Maklaya ul., Moscow 117198, Russia

\end{center}

\vspace{15pt}

\small\noi

\begin{abstract}

Black hole  generalized $p$-brane  solutions  for a wide class of
intersection rules are obtained. The solutions are defined on a
manifold that contains a product of $n - 1$ Ricci-flat internal
spaces. They are defined up to  a set of functions $H_s$ obeying
non-linear differential equations equivalent to Toda-type
equations with certain boundary conditions imposed. A conjecture
on polynomial structure of governing functions $H_s$ for
intersections related to semisimple Lie algebras is suggested.
This conjecture is proved for Lie algebras: $A_m$, $C_{m+1}$, $m
\geq 1$. For simple Lie algebras the powers of polynomials
coincide with the components  of twice the dual Weyl vector in the
basis of simple coroots. The coefficients of polynomials depend
upon the extremality parameter $\mu >0$. In the extremal case $\mu
= 0$ such polynomials were considered previously by H. L\"u, J.
Maharana, S. Mukherji  and C.N. Pope. Explicit formulas for
$A_2$-solution are obtained.  Two examples of $A_2$-dyon
solutions, i.e. dyon in $D = 11$ supergravity with $M2$ and $M5$
branes intersecting at a point and  Kaluza-Klein dyon, are
considered.

\end{abstract}

\hspace{1cm}*PACS numbers:
04.20.Jb, 04.50.+h, 04.70.Bw, 02.20.Sv, 02.30.Hq.

\pagebreak

\normalsize

\section{Introduction}

 At present there exists an interest to the so-called $M$-theory
(see, for example, \cite{M-th1}-\cite{M-th2}).
This theory is ``supermembrane'' analogue of
superstring models \cite{GSW} in $D=11$. The low-energy limit of
$M$-theory after a dimensional reduction leads to models governed by a
Lagrangian containing a metric, fields of  forms and scalar fields.
These models contain a large variety of the so-called
$p$-brane solutions (see \cite{St}-\cite{IK} and references
therein).

In \cite{IMC} it was shown that after
the dimensional reduction on the
manifold $M_0\times M_1\times\dots\times M_n$  when the composite
$p$-brane ansatz for fields of forms
is considered the problem is reduced to the gravitating
self-interacting $\sigma$-model with certain constraints imposed. (For
electric $p$-branes see also \cite{IM0,IM,IMR}.) This representation
may be considered as a tool for obtaining different solutions
with intersecting $p$-branes. In \cite{IMC,IMR,IMBl,IKM,GrI}
the Majumdar-Papapetrou type solutions (see \cite{MP})
were obtained (for non-composite
case see \cite{IM0,IM}). These solutions corresponding to Ricci-flat
factor-spaces $(M_i,g^i)$, ($g^i$ is metric on $M_i$)
$i=1,\dots,n$, were also
generalized to the case of Einstein internal spaces \cite{IMC}.  Earlier
some special classes of these solutions were considered in
\cite{Ts1,PT,GKT,AR,AEH,AIR}. The obtained solutions take place, when
certain (block-)orthogonality relations (on couplings parameters,
dimensions of "branes", total dimension) are imposed. In this situation a
class of cosmological and spherically-symmetric solutions was obtained
\cite{IMJ,Br2,IMJ2}. Special cases were also considered in
\cite{LPX,BGIM,GrIM,BKR}.  The solutions with the horizon were studied
in details in \cite{CT,AIV,Oh,IMJ,BIM,IMBl,Br2,CIM}.

In models under consideration  there exists a large variety of
Toda-chain solutions,  when certain intersection rules are satisfied
\cite{IMJ}.  Cosmological and spherically symmetric solutions with
$p$-branes and $n$ internal spaces  related to $A_m$ Toda chains
were previously  considered in   \cite{LPX,LMPX} and   \cite{GM1,GM2}.
Recently in \cite{IK} a family of $p$-brane  solutions
depending on  one harmonic function with nearly arbitrary
(up to some restrictions) intersection rules  were obtained.
These solutions are defined up to
solutions of Laplace and Toda-type equations and correspond to
null-geodesics of the sigma-model target-space metric.

Here we consider a family of spherically-symmetric and
cosmological type solutions from \cite{IK}
(see Sect. 2) and single out a new subclass of black-hole configurations
related to Toda-type equations with
certain asymptotical conditions  imposed (Sect. 3).
These black hole solutions are
governed by functions $H_s(z) > 0$ defined
on the interval $(0, (2 \mu)^{-1})$ ($\mu > 0$)
and obeying a set of second order non-linear differential equations
\beq{0.1}
 \frac{d}{dz} \left( \frac{(1 - 2\mu z)}{H_s} \frac{d}{dz} H_s \right) =
 \bar B_s \prod_{s' \in S}  H_{s'}^{- A_{s s'}},
\eeq
with  the following boundary conditions imposed:
\begin{eqnarray}
\label{0.2}
&&{\bf (i)} \quad H_{s}((2\mu)^{-1} -0) = H_{s0} \in (0, + \infty);
\\ \label{0.3}
&&{\bf (ii)}  \quad H_{s}(+ 0) = 1,
\end{eqnarray}
$s \in S$.  In (\ref{0.1})
$\bar B_s > 0$, $s \in S$, and  $(A_{s s'})$ is a "quasi-Cartan" matrix
($A_{ss} = 2$, $s \in S$)
coinciding with the Cartan one when intersections are related to Lie
algebras. Equations (\ref{0.1}) are equivalent to Toda-type equations
(see Sect. 2).

For positively defined scalar field metric $(h_{\alpha \beta})$ all
$p$-branes in this solution should contain a time manifold
(see Proposition 1 from Sect. 3).
This agrees with Theorem  3 from Ref. \cite{Br2}
(for orthogonal case, see also \cite{BIM}).

In Sect. 4 we suggest a hypothesis: the functions $H_s$
are polynomials when intersection rules correspond to
semisimple Lie algebras.  This hypothesis ({\bf Conjecture } )
is proved for Lie algebras: $A_m$, $C_{m+1}$, $m = 1,2, \ldots$.
It is also confirmed by special  black-hole ``block orthogonal''
solutions considered earlier in \cite{Br,IMBl,CIM,IMJ2}.
An analogue of this conjecture for extremal black holes
was considered earlier in \cite{LMMP}.  In Sect. 5  explicit formulas for
the solution corresponding to the algebra $A_2$ are obtained.  These
formulas are illustrated by two examples of $A_2$-dyon solutions: a dyon
in $D = 11$ supergravity (with $M2$ and $M5$ branes intersecting at a point)
and Kaluza-Klein dyon.


\section{The model and Toda-type solutions}

We consider a  model governed by
the action \cite{IMC}
\ber{1.1}
S=\int d^Dx \sqrt{|g|}\biggl\{R[g]-h_{\alpha\beta}g^{MN}\p_M\varphi^\alpha
\p_N\varphi^\beta-\sum_{a\in\tri}\frac{\theta_a}{n_a!}
\exp[2\lambda_a(\varphi)](F^a)^2\biggr\}
\eer
where $g=g_{MN}(x)dx^M\otimes dx^N$ is a metric,
$\varphi=(\varphi^\alpha)\in\R^l$ is a vector of scalar fields,
$(h_{\alpha\beta})$ is a  constant symmetric
non-degenerate $l\times l$ matrix $(l\in \N)$,
$\theta_a=\pm1$,
\beq{1.2a}
F^a =    dA^a
=  \frac{1}{n_a!} F^a_{M_1 \ldots M_{n_a}}
dz^{M_1} \wedge \ldots \wedge dz^{M_{n_a}}
\eeq
is a $n_a$-form ($n_a\ge1$), $\lambda_a$ is a
1-form on $\R^l$: $\lambda_a(\varphi)=\lambda_{\alpha a}\varphi^\alpha$,
$a\in\tri$, $\alpha=1,\dots,l$.
In (\ref{1.1})
we denote $|g| =   |\det (g_{MN})|$,
\beq{1.3a}
(F^a)^2_g  =
F^a_{M_1 \ldots M_{n_a}} F^a_{N_1 \ldots N_{n_a}}
g^{M_1 N_1} \ldots g^{M_{n_a} N_{n_a}},
\eeq
$a \in \tri$. Here $\tri$ is some finite set.

Let us consider a family of one-variable sector
solutions to field equations corresponding to the action
(\ref{1.1}) and depending upon one variable $u$
\cite{IK}. These solutions are defined on the manifold
\beq{1.2}
M =    (u_{-}, u_{+})  \times
M_1  \times M_2 \times  \ldots \times M_{n},
\eeq
where $(u_{-}, u_{+})$  is  an interval belonging to $\R$.
The solutions read \cite{IK}
\bear{1.3}
g= \biggl(\prod_{s \in S} [f_s(u)]^{2 d(I_s) h_s/(D-2)} \biggr)
\biggl\{[f_1(u)]^{2d_1/(1-d_1)}\exp(2c^1u + 2 \bar c^1)\\ \nn
\times[w du \otimes du+ f_1^2(u)g^1] +
\sum_{i = 2}^{n} \Bigl(\prod_{s\in S}
[f_s(u)]^{- 2 h_s  \delta_{i I_s} } \Bigr)
\exp(2c^i u+ 2 \bar c^i) g^i\biggr\}, \\ \label{1.4}
\exp(\varphi^\alpha) =
\left( \prod_{s\in S} f_s^{h_s \chi_s \lambda_{a_s}^\alpha} \right)
\exp(c^\alpha u + \bar c^\alpha), \\ \label{1.5}
F^a= \sum_{s \in S} \delta^a_{a_s} {\cal F}^{s},
\ear
$\alpha=1,\dots,l$.
In  (\ref{1.3})  $w = \pm 1$,
$g^i=g_{m_i n_i}^i(y_i) dy_i^{m_i}\otimes dy_i^{n_i}$
is a Ricci-flat  metric on $M_{i}$, $i=  2,\ldots,n$,
the space  $(g^1, M_{1})$ is an Einstein space
of non-zero curvature:
\beq{1.5a}
R_{m n}[g^1 ] = \xi^1 g^1,
\eeq
$\xi^1 \neq 0$, and
\beq{1.11}
\delta_{iI}=  \sum_{j\in I} \delta_{ij}
\eeq
is the indicator of $i$ belonging
to $I$: $\delta_{iI}=  1$ for $i\in I$ and $\delta_{iI}=  0$ otherwise.

The  $p$-brane  set  $S$ is by definition
\ber{1.6}
S=  S_e \cup S_m, \quad
S_v=  \cup_{a\in\tri}\{a\}\times\{v\}\times\Omega_{a,v},
\eer
$v=  e,m$ and $\Omega_{a,e}, \Omega_{a,m} \subset \Omega$,
where $\Omega =   \Omega(n)$  is the set of all non-empty
subsets of $\{ 2, \ldots,n \}$. Hence
all $p$-branes do not ``live'' in  $M_1$.

Any $p$-brane index $s \in S$ has the form
\ber{1.7}
s =   (a_s,v_s, I_s),
\eer
where
$a_s \in \tri$, $v_s =  e,m$ and $I_s \in \Omega_{a_s,v_s}$.
The sets $S_e$ and $S_m$ define electric and magnetic $p$-branes
correspondingly. In
(\ref{1.4})
\ber{1.8}
\chi_s  =   +1, -1
\eer
for $s \in S_e, S_m$ respectively.
In (\ref{1.5})  forms
\beq{1.9}
{\cal F}^s= Q_s
\left( \prod_{s' \in S}  f_{s'}^{- A_{s s'}} \right) du \wedge\tau(I_s),
\eeq
$s\in S_e$, correspond to electric $p$-branes and
forms
\beq{1.10}
{\cal F}^s= Q_s \tau(\bar I_s),
\eeq
correspond to magnetic $p$-branes; $Q_s \neq 0$, $s \in S$.
In (\ref{1.10})  and in what follows
\beq{1.13a}
\bar I\equiv\{1,\ldots,n\}\setminus I.
\eeq

All the  manifolds $M_{i}$, $i > 1$, are assumed to be oriented and
connected and  the volume $d_i$-forms
\beq{1.12}
\tau_i  \equiv \sqrt{|g^i(y_i)|}
\ dy_i^{1} \wedge \ldots \wedge dy_i^{d_i},
\eeq
are well--defined for all $i=  1,\ldots,n$.
Here $d_{i} =   {\rm dim} M_{i}$, $i =   1, \ldots, n$
(in spherically symmetric case $M_1 = S^{d_1}$), $d_1 > 1$,
$D =   1 + \sum_{i =   1}^{n} d_{i}$, and for any
 $I =   \{ i_1, \ldots, i_k \} \in \Omega$, $i_1 < \ldots < i_k$,
we denote
\ber{1.13}
\tau(I) \equiv \tau_{i_1}  \wedge \ldots \wedge \tau_{i_k},
\\ \label{1.14}
M_{I} \equiv M_{i_1}  \times  \ldots \times M_{i_k}, \\
\label{1.15}
d(I) \equiv {\rm dim } M_I =  \sum_{i \in I} d_i.
\eer

The parameters  $h_s$ appearing in the solution
satisfy the relations
\beq{1.16}
h_s = K_s^{-1}, \qquad  K_s = B_{s s},
\eeq
where
\ber{1.17}
B_{ss'} \equiv
d(I_s\cap I_{s'})+\frac{d(I_s)d(I_{s'})}{2-D}+
\chi_s\chi_{s'}\lambda_{\alpha a_s}\lambda_{\beta a_{s'}}
h^{\alpha\beta},
\eer
$s, s' \in S$, with $(h^{\alpha\beta})=(h_{\alpha\beta})^{-1}$.
Here we assume that
\beq{1.17a}
({\bf i}) \qquad B_{ss} \neq 0,
\eeq
for all $s \in S$, and
\beq{1.18b}
({\bf ii}) \qquad {\rm det}(B_{s s'}) \neq 0,
\eeq
i.e. the matrix $(B_{ss'})$ is a non-degenerate one. In (\ref{1.9})
another non-degenerate matrix (``a quasi-Cartan'' matrix)
appears
\beq{1.18}
(A_{ss'}) = \left( 2 B_{s s'}/B_{s' s'} \right).
\eeq
Here  some ordering in $S$ is assumed.

This matrix also appears in  the relations for
\beq{1.19}
f_s = \exp( - q^s),
\eeq
where $(q^s) = (q^s(u))$ is a solution to Toda-type equations
\beq{1.20}
\ddot{q^s} = -  B_s \exp( \sum_{s' \in S} A_{s s'} q^{s'} ),
\eeq
with
\beq{1.21}
 B_s = 2 K_s A_s,  \quad  A_s =  \frac12  \eps_s Q_s^2,
\eeq
$s \in S$. Here
\beq{1.22}
\eps_s=(-\eps[g])^{(1-\chi_s)/2}\eps(I_s) \theta_{a_s},
\eeq
$s\in S$, $\eps[g]\equiv\sign\det(g_{MN})$. More explicitly
(\ref{1.22}) reads: $\eps_s=\eps(I_s) \theta_{a_s}$ for
$v_s = e$ and $\eps_s=-\eps[g] \eps(I_s) \theta_{a_s}$, for
$v_s = m$.

In (\ref{1.3})
\bear{1.23}
f_1(u) =R \sh(\sqrt{C_1}u), \ C_1>0, \ \xi_1 w>0;
\\ \label{1.24}
R \sin(\sqrt{|C_1|}u), \ C_1<0, \  \xi_1 w>0;   \\ \label{1.25}
R \ch(\sqrt{C_1}u),  \ C_1>0, \ \xi_1w <0; \\ \label{1.26}
\left|\xi_1(d_1-1)\right|^{1/2} u, \ C_1=0,  \ \xi_1w>0,
\ear
where  $C_1$ is constant and $R =  |\xi_1(d_1-1)/C_1|^{1/2}$.

Vectors $c=(c^A)= (c^i, c^\alpha)$ and
$\bar c=(\bar c^A)$ satisfy the linear constraints
\bear{1.27}
U^s(c)= \sum_{i \in I_s}d_ic^i-\chi_s\lambda_{a_s\alpha}c^\alpha=0,
\\ \label{1.28}
U^s(\bar c)=  \sum_{i\in I_s}d_i\bar c^i-
\chi_s\lambda_{a_s\alpha}\bar c^\alpha=0,
\ear
$s\in S$,
\bear{1.29}
U^1( c) =  -c^1+\sum_{j=1}^n d_j c^j=0, \\ \label{1.30}
U^1(\bar c) =
-\bar c^1+\sum_{j=1}^nd_j\bar c^j=0,
\ear
and
\beq{1.30a}
C_1 \frac{d_1}{d_1-1}= 2 E_{TL} +
h_{\alpha\beta}c^\alpha c^\beta+ \sum_{i=2}^nd_i(c^i)^2+
\frac1{d_1-1}\left(\sum_{i=2}^nd_ic^i\right)^2,
\eeq
where
\beq{1.31}
E_{TL} = \frac{1}{4}  \sum_{s,s' \in S} h_s
A_{s s'} \dot{q^s} \dot{q^{s'}}
  + \sum_{s \in S} A_s  \exp( \sum_{s' \in S} A_{s s'} q^{s'} ),
\eeq
is an integration constant (energy) for the solutions from
(\ref{1.20}).

We note that the eqs. (\ref{1.20}) correspond to the
Toda-type Lagrangian
\beq{1.31a}
L_{TL} = \frac{1}{4}  \sum_{s,s' \in S}
h_s  A_{s s'} \dot{q^s}\dot{q^{s'}}
-  \sum_{s \in S} A_s  \exp( \sum_{s' \in S} A_{s s'} q^{s'} ).
\eeq

{\bf Remark 1.}
{\em Here we identify notations  for $g^{i}$  and  $\hat{g}^{i}$, where
$\hat{g}^{i} = p_{i}^{*} g^{i}$ is the
pullback of the metric $g^{i}$  to the manifold  $M$ by the
canonical projection: $p_{i} : M \rightarrow  M_{i}$, $i = 1,
\ldots, n$. An analogous agreement will be also kept for volume forms etc.}

Due to (\ref{1.9}) and  (\ref{1.10}), the dimension of
$p$-brane worldsheet $d(I_s)$ is defined by
\ber{1.16a}
d(I_s)=  n_{a_s}-1, \quad d(I_s)=   D- n_{a_s} -1,
\eer
for $s \in S_e, S_m$ respectively.
For a $p$-brane: $p =   p_s =   d(I_s)-1$.

The solutions are valid if the following  restrictions on the sets
$\Omega_{a,v}$ are imposed.
These restrictions guarantee the block-diagonal structure
of the stress-energy tensor, like for the metric, and the existence of
$\sigma$-model representation \cite{IMC} (see also \cite{AR}).
We denote $w_1\equiv\{i|i\in \{2,\dots,n\},\quad d_i=1\}$, and
$n_1=|w_1|$ (i.e. $n_1$ is the number of 1-dimensional spaces among
$M_i$, $i=1,\dots,n$).

{\bf Restriction 1.} {\em Let 1a) $n_1\le1$ or 1b) $n_1\ge2$ and for
any $a\in\tri$, $v\in\{e,m\}$, $i,j\in w_1$, $i<j$, there are no
$I,J\in\Omega_{a,v}$ such that $i\in I$, $j\in J$ and $I\setminus\{i\}=
J\setminus\{j\}$.}

{\bf Restriction 2.} {\em Let 2a) $n_1=0$ or
2b) $n_1\ge1$ and for any $a\in\tri$, $i\in w_1$ there are no
$I\in\Omega_{a,m}$, $J\in\Omega_{a,e}$ such that $\bar I=\{i\}\sqcup J$.}

These restrictions are  satisfied in the non-composite case
\cite{IM0,IM}:  $|\Omega_{a,e}| + |\Omega_{a,m}| = 1$,
(i.e when there are no two  $p$-branes with the same color index $a$,
$a\in\tri$.) Restriction 1 and 2 forbid certain intersections of two
$p$-branes with the same color index for  $n_1 \geq 2$ and  $n_1 \geq 1$
respectively.  Restriction 2 is satisfied identically if all
$p$-branes contain a common manifold $M_j$ (say, time manifold).

This solution describes a set of charged (by forms) overlapping
$p$-branes ($p_s=d(I_s)-1$, $s \in S$) ``living'' on submanifolds
of $M_2 \times \dots \times M_n$.

\subsection{$U^s$-vectors and scalar products}

Here we consider a minisuperspace
covariant form of constraints
and corresponding scalar products
that will be used in the next section.
The linear constraints (\ref{1.27})-(\ref{1.30})
may be written in the following form
\beq{1.32}
U^r(c)= U^r_A c^A= 0, \qquad U^r(\bar c)= U^r_A \bar c^A= 0,
\eeq
$r = s,1$, where
\ber{1.33}
(U_A^s)=(d_i\delta_{iI_s},-\chi_s\lambda_{\alpha a_s}),
\eer
$s=(a_s,v_s,I_s) \in S$, and
\ber{1.34}
(U_A^1)=(- \delta^1_i + d_i, 0),
\eer
$A = (i, \alpha)$.

The quadratic constraint (\ref{1.30a}) reads
\beq{1.30b}
E=E_1+ E_{TL}+ \frac12 \hat G_{AB} c^A c^B = 0,
\eeq
where $C_1=2E_1(U^1,U^1)$,
\beq{1.35}
(U^1,U^1) = 1/d_1 - 1,
\eeq
($d_1 > 1$) and
\ber{1.36}
(\hat G_{AB})=\barr{cc}
G_{ij}& 0\\
0& h_{\alpha\beta}
\earr,
\eer
is the target space metric with
\ber{1.37}
G_{ij}= d_i \delta_{ij} - d_i d_j,
\eer
$i,j = 1, \ldots, n$.
In (\ref{1.35})  a scalar product appears

\ber{1.38}
(U,U')=\hat G^{AB}U_AU'_B,
\eer

where $U=U_Az^A$, $U' = U'_A z^A$ are linear functions on
$\R^{n+l}$, and  $(\hat G^{AB})=(\hat G_{AB})^{-1}$. The scalar
products (\ref{1.38}) for co-vectors $U^s$ from $(\ref{1.33})$
were calculated in \cite{IMC}

\beq{1.39}
(U^s,U^{s'})= B_{s s'},
\eeq
$s, s' \in S$ (see
(\ref{1.17})). It follows from (\ref{1.18b}) and  (\ref{1.39})
that the vectors $U^s$, $s \in S$, are  linearly independent.
Hence, the number of the vectors $U^s$ should not exceed the
dimension of the dual space $({\bf R}^{n+ l})^{*}$, i.e.
 \beq{1.40c}
 |S| \leq n+ l.
 \eeq
We also get \cite{IMC}
 \beq{1.39a}
 (U^s,U^{1})= 0,
 \eeq
for all $s \in S$. This relation takes place,
since all $p$-branes do not live in $M_1$: $I_s \in \{2,\ldots,n
 \}$.

{\bf Intersection rules.}
From  (\ref{1.16}), (\ref{1.17}) and (\ref{1.18})  we get
the  intersection rules  corresponding
to the quasi-Cartan matrix $(A_{s s'})$ \cite{IMJ}
\beq{1.40}
d(I_s \cap I_{s'})= \frac{d(I_s)d(I_{s'})}{D-2}-
\chi_s\chi_{s'}\lambda_{a_s}\cdot\lambda_{a_{s'}} + \frac12 K_{s'} A_{s s'},
\eeq
where $\lambda_{a_s}\cdot\lambda_{a_{s'}} =
\lambda_{\alpha a_s}\lambda_{\beta a_{s'}} h^{\alpha\beta}$,
$s, s' \in S$.

The contravariant components $U^{rA}= \hat G^{AB} U^r_B$ reads
\cite{IMC,IMJ}
\beq{1.41}
U^{si}= G^{ij}U_j^s=
\delta_{iI_s}-\frac{d(I_s)}{D-2}, \quad U^{s\alpha}= - \chi_s
\lambda_{a_s}^\alpha,
\eeq
\beq{1.42}
U^{1i}=-\frac{\delta_1^i}{d_1},
\quad U^{1\alpha}=0,
\eeq
$s \in S$.
Here (as in \cite{IMZ})
\beq{1.43}
G^{ij}=\frac{\delta^{ij}}{d_i}+\frac1{2-D},
\eeq
$i,j=1,\dots,n$, are the components of the matrix inverse to
$(G_{ij})$ from (\ref{1.37}). The contravariant components
(\ref{1.41}) and  (\ref{1.42}) occur as powers in relations
for the metric and scalar fields in (\ref{1.3}) and  (\ref{1.4}).

We note that the solution under consideration for the special case
of the $A_m$ Toda chain was obtained earlier  in \cite{GM1}.
Special $A_1 \oplus \dots \oplus A_1$ Toda case, when vectors
$U^s$ are mutually orthogonal, was considered earlier in \cite{IMJ}
(for non-composite case see also \cite{BGIM,GrIM,BIM}).
For a (general) block-orthogonal set of vectors $U^s$
special solutions were considered in \cite{Br,IMJ2}.

\section{ Black holes solutions}

\subsection{The choice of parameters }

Here we consider the spherically symmetric case:
\beq{2.1}
w = 1, \quad M_1 = S^{d_1}, \quad g^1 = d \Omega^2_{d_1},
\eeq
where $d \Omega^2_{d_1}$ is the canonical metric on a unit sphere
$S^{d_1}$, $d_1 \geq 2$. In this case $\xi^1 = d_1 -1$.
We also assume that
\beq{2.2}
M_2 = \R, \qquad g^2 = - dt \otimes dt,
\eeq
i.e.  $M_2$ is a time manifold.

We put $C_1 \geq 0$.
In this case relations (\ref{1.23})-(\ref{1.26}) read
\bear{1.23s}
f_1(u) = \bar d \frac{\sh(\sqrt{C_1}u)}{\sqrt{C_1}},
\ C_1>0,
\\ \label{1.26s}
\bar du, \ C_1=0.
\ear
Here and in what follows
\beq{2.3}
\bar d = d_1-1.
\eeq

Let us consider the null-geodesic equations
for the light ``moving'' in the radial direction
(following from $ds^2 =0$):
\bear{2.4}
 \frac{dt}{du} = \pm \Phi, \\ \label{2.4a}
 \Phi =  f_1^{d_1/(1-d_1)} e^{(c^1 - c^2) u +  \bar c^1 - \bar c^2}
 \prod_{s\in S} f_s^{h_s  \delta_{2 I_s}},
\ear
equivalent to
\beq{2.5}
 t - t_0 = \pm \int_{u_0}^{u} d \bar u  \Phi(\bar u),
\eeq
where $t_0, u_0$ are constants.

Let us consider   solutions
(defined on some interval $[u_0, +\infty)$) with a  horizon
at $u = + \infty$ satisfying
\beq{2.6}
  \int_{u_0}^{ + \infty} d  u  \Phi( u) = + \infty.
\eeq

Here we restrict ourselves to  solutions with
$C_1 > 0$ and linear asymptotics at infinity
\beq{2.7}
q^s = - \beta^s u + \bar \beta^s  + o(1),
\eeq
$u \to +\infty$, where $\beta^s, \bar \beta^s$ are
constants, $s \in S$. This relation gives us an
asymptotical solution to  Toda type eqs. (\ref{1.20}) if
\beq{2.8}
\sum_{s' \in S} A_{s s'} \beta^{s'} > 0,
\eeq
for all $s \in S$. In this case the energy  (\ref{1.31})
reads
\beq{2.9}
E_{TL} = \frac{1}{4}  \sum_{s,s' \in S}
h_s A_{s s'}  \beta^s \beta^{s'}.
\eeq

{\bf Remark 2.} {\em  For positive-definite matrices
$(h_s A_{s s'})$ and  $(h_{\alpha\beta})$ we get from
(\ref{1.30a}) and (\ref{2.9}):
$E_{TL} \geq 0$,  $C_{1} \geq 0$.
(For the extremal case $E_{TL} = C_{1} = 0$ see Sect. 7.)
According to  Lemma 2 from \cite{Br2}
black hole solutions can only exist for
$C_{1} \geq 0$ and the horizon is then at
$u = \infty$.}

For  the function  (\ref{2.4a}) we get
\beq{2.10}
\Phi(u) \sim \Phi_0 e^{\beta u}, \quad u \to +\infty,
\eeq
where  $\Phi_0 \neq 0$ is  constant,
\beq{2.11}
 \beta = c^1 - c^2 + \sqrt{C_1} h_1
+ \sum_{s \in S} \beta_s h_s  \delta_{2 I_s},
\eeq
and
\beq{2.12}
  h_1 = (U^1, U^1)^{-1} = \frac{d_1}{1 - d_1}.
\eeq
Horizon at $u = + \infty$  takes place if and only if
\beq{2.12a}
  \beta \geq 0.
\eeq
Let us introduce dimensionless parameters
\beq{2.13}
b^s = \beta^s / \sqrt{C_1}, \qquad
b^A =  c^A / \sqrt{C_1},
\eeq
where $s \in S$, $A = (i, \alpha)$,
$C_1 > 0$.

Thus, a horizon at $u = + \infty$
corresponds to a point $b = (b^s,b^A) \in  \R^{|S| + n +l}$
satisfying the relations following from
(\ref{1.32}), (\ref{1.30b}), (\ref{2.8}), (\ref{2.9})
and (\ref{2.11})-(\ref{2.13}):
\bear{2.14}
U^r_A b^A= 0,   \qquad r = s,1; \ s \in S, \\ \label{2.15}
\frac{1}{2}  \sum_{s,s' \in S} h_s A_{s s'}  b^s b^{s'}
+ \hat G_{AB} b^A b^B = |h_1|, \\
\label{2.16}
\sum_{s' \in S} A_{s s'} b^{s'} > 0,  \\ \label{2.17}
 f(b) \equiv
 b^1 - b^2  + \sum_{s\in S} b_s h_s  \delta_{2 I_s} \geq |h_1|.
\ear

{\bf Proposition 1.} {\em Let  matrix $(h_{\alpha\beta})$
be positively defined. Then the point $b = (b^s,b^A)$ satisfying
relations (\ref{2.14})-(\ref{2.17}) exists only if
\beq{2.18}
2 \in I_s, \quad \forall s \in S,
\eeq
(i. e. all p-branes have a common time direction $t$)
and is unique: $b = b_0$, where
\bear{2.19}
b_0^A  = - \delta^{A}_{2} + h_1 U^{1 A}  +
\sum_{s\in S}  h_s b_0^s U^{s A},  \\ \label{2.20}
b_0^s = 2 \sum_{s' \in S} A^{s s'},
\ear
where $s \in S$, $A = (i, \alpha)$,
and the matrix $(A^{s s'})$ is inverse to the matrix
$(A_{s s'}) = ( 2 (U^s, U^{s'})/ (U^{s'}, U^{s'}))$. }

{\bf Proof.} Let ${\cal E}$ be a manifold described
by relations (\ref{2.14})-(\ref{2.15}). This manifold is an ellipsoid.
Indeed, due to positively definiteness of $(h_{\alpha\beta})$
the matrix $\hat G_{AB}$ has a signature $(-,+, \ldots,+)$,
since the matrix $(G_{ij})$ from (\ref{1.37}) has a signature
$(-,+, \ldots,+)$ \cite{IMZ}. Due to relations
$(U^1,U^1) < 0$,  $(U^1,U^s) = 0$, $(U^s,U^s) \neq  0$ for all $s \in S$,
and (\ref{1.18b}) the matrices $(B_{s s'})$ and $(A_{s s'})$ are positively
defined and all $h_s > 0$, $s \in S$. Then, the quadratic form
in (\ref{2.15}) has a pseudo-Euclidean signature.
Due to $(U^1,U^1) < 0$ the intersection of the
hyperboloid (\ref{2.15}) with the (multidimensional)
plane $U^1_A z^A = 0$ gives us an ellipsoid. Its intersection
with the planes $U^s_A z^A = 0$, $s \in S$, give us to an ellipsoid,
coinciding with ${\cal E}$.

Let us consider  a function $f_{|} : {\cal E} \rightarrow \R$
that is a restriction of the linear function
(\ref{2.17}) on ${\cal E}$.
Let $b_{*} \in {\cal E}$ be a point of maximum
of $f_{|}$.
Using the conditional extremum method
and the  fact that ${\cal E}$ is  ellipsoid
we  prove that

that
\bear{2.21}
b_{*}^A = - \delta^{A}_{2} + h_1 U^{1 A} +
\sum_{s\in S}  h_s b_{*}^s U^{s A},  \\ \label{2.21a}
b_{*}^s = 2 \sum_{s' \in S} A^{s s'} \delta_{2 I_{s'}},
\ear
$s \in S$, $A = (i, \alpha)$.
Let us consider the function
\bear{2.22}
 \bar f(b, \lambda) \equiv  f(b)
 - \lambda_1 U^1_A b^A
 - \sum_{s \in S} \lambda_s U^s_A b^A
 - \lambda_0
\left(\sum_{s,s' \in S} \frac{h_s}{2} A_{s s'} b^s b^{s'}
+ \hat G_{AB} b^A b^B + h_1 \right),
\ear
where
$\lambda = (\lambda_0, \lambda_1, \lambda_s)$ is a vector of
Lagrange multipliers. The  points of extremum for the
function  $\bar f$ from (\ref{2.22}) have the
form $(\lambda_0  b_{*}, \lambda)$ with $b_{*}$ from (\ref{2.21}) and
\bear{2.23}
  \lambda_0 = \pm 1, \quad   \lambda_1 = 1/(d_1 -1), \quad
  \lambda_s = - 2 \sum_{s' \in S} h_s A^{s s'} \delta_{2 I_{s'}},
\ear
$s \in S$. Then, the  points $b_{*}$ and $ - b_{*}$
are the points of maximum
and minimum, respectively, for the function $f_{|}$ defined on the
ellipsoid  ${\cal E}$. Since $f(b_{*}) = |h_1|$, the only point
satisfying the restriction $f(b) \geq |h_1|$ is $b = b_{*}$.
From (\ref{2.16}) we get
\bear{2.24}
\sum_{s' \in S} A_{s s'} b^{s'}= 2 \delta_{2 I_{s}} > 0
\Longleftrightarrow  2 \in I_s,
\ear
for all $s \in S$. The proposition is proved.

We introduce a new radial variable $R = R(u)$ by relations
\bear{2.28}
\exp( - 2\bar{\mu} u) = 1 - \frac{2\mu}{R^{\bar{d}}} = F,
\qquad \bar{\mu} = \sqrt{C_1}, \quad
\mu = \bar{\mu}/ \bar{d} >0,
\ear
$u > 0$, $R^{\bar d} > 2\mu$ ($\bar d = d_1 -1$).
We put
\bear{2.27}
\bar{c}^A = 0,
\\     \label{2.27f}
q^s(0) = 0.
\ear
$A = (i, \alpha)$, $s \in S$.
These relations guarantee the asymptotical flatness
(for $R \to +\infty$) of the $(2+d_1)$-dimensional section of the metric.

Let us denote
\beq{2.28a}
H_s = f_s e^{- \bar{\mu} b^s_0 u },
\eeq
$s \in S$.
Then,  solutions (\ref{1.3})-(\ref{1.5}) may be written as follows
\bear{2.30}
g= \Bigl(\prod_{s \in S} H_s^{2 h_s d(I_s)/(D-2)} \Bigr)
\biggl\{ F^{-1} dR \otimes dR
+ R^2  d \Omega^2_{d_1}  \\ \nn
-  \Bigl(\prod_{s \in S} H_s^{-2 h_s} \Bigr) F  dt \otimes dt
+ \sum_{i = 3}^{n} \Bigl(\prod_{s\in S}
  H_s^{-2 h_s \delta_{iI_s}} \Bigr) g^i  \biggr\},
\\  \label{2.31}
\exp(\varphi^\alpha)=
\prod_{s\in S} H_s^{h_s \chi_s \lambda_{a_s}^\alpha},
\\  \label{2.32a}
F^a= \sum_{s \in S} \delta^a_{a_s} {\cal F}^{s},
\ear
where
\beq{2.32}
{\cal F}^s= - \frac{Q_s}{R^{d_1}}
\left( \prod_{s' \in S}  H_{s'}^{- A_{s s'}} \right) dR \wedge\tau(I_s),
\eeq
$s\in S_e$,
\beq{2.33}
{\cal F}^s= Q_s \tau(\bar I_s),
\eeq
$s\in S_m$.
Here $Q_s \neq 0$, $h_s =K_s^{-1}$; parameters $K_s \neq 0$ and
the non-degenerate
matrix $(A_{s s'})$ are defined by  relations (\ref{1.40})
and $(A_{s s}) = 2$, $s \in S$.

Functions $H_s > 0$ obey the equations
\beq{2.34}
 R^{d_1} \frac{d}{dR} \left( R^{d_1}
\frac{F}{H_s}   \frac{d H_s}{dR} \right) = B_s
\prod_{s' \in S}  H_{s'}^{- A_{s s'}},
\eeq
$s \in S$, where $B_s \neq 0$ are defined
in (\ref{1.21}) and (\ref{1.22}).
These equations follow from Toda-type equations (\ref{1.20}) and
the definition   (\ref{2.28}) and   (\ref{2.28a}).

It follows from (\ref{2.7}), (\ref{2.13}) and  (\ref{2.28a})
that there exist finite limits
\beq{2.35a}
H_s  \to H_{s0} \neq 0,
\eeq
for $R^{\bar d} \to 2\mu$, $s \in S$.
We note, that in this case the metric (\ref{2.30})
does really have a horizon at  $R^{\bar{d}} =   2 \mu$.

From (\ref{2.27f})  we get. \beq{2.35} H_s (R = +\infty) = 1, \eeq
$s \in S$.

The metric (\ref{2.30}) has a regular horizon at
$R^{\bar{d}} =   2 \mu$.
The Hawking temperature corresponding to
the solution is (see also \cite{Oh,BIM} for orthogonal case)
found to be
 \beq{2.36}
 T_H=   \frac{\bar{d}}{4 \pi (2 \mu)^{1/\bar{d}}} \prod_{s \in S}
 H_{s0}^{- h_s},
\eeq
where $H_{s0}$ are defined in (\ref{2.35a}).

The boundary conditions (\ref{2.35a}) and  (\ref{2.35})
play a crucial role here, since they
single out, generally speaking, only few solutions
to eqs. (\ref{2.34}).

Moreover for some values of parameters $\mu = \bar \mu / \bar d$, $\eps_s$
and $Q_s^2$ the solutions to eqs. (\ref{2.34})-(\ref{2.35})
do not exist. Indeed, from (\ref{1.21}), (\ref{1.31}),
(\ref{2.9}), (\ref{2.13}), (\ref{2.20}), (\ref{2.28}) and
(\ref{2.27f} we get
\beq{2.35b}
E_{TL} = \bar \mu^2  \sum_{s,s' \in S} h_s A^{ss'} =
\frac{1}{4}  \sum_{s,s' \in S} h_s A_{s s'}
\dot{q^s}(0) \dot{q^{s'}}(0) + \sum_{s \in S}
\frac12  \eps_s Q_s^2.
\eeq
Let the matrix $(h_s A_{s s'})$ be positive-definite
(in this case matrix $(B_{s s'})$ is positive-definite
too and all $h_s > 0$). Then $E_{TL} > 0$ and
\beq{2.35c}
\bar \mu^2  \sum_{s,s' \in S} h_s A^{ss'} \geq
\sum_{s \in S}   \frac12  \eps_s Q_s^2.
\eeq
If the parameters  obey the relation
\beq{2.35d}
0 < \bar \mu^2  \sum_{s,s' \in S} h_s A^{ss'} <
\sum_{s \in S}  \frac12  \eps_s Q_s^2,
\eeq
e.g. for $\eps_s = + 1$ and big enough $Q_s^2$, the
solution under consideration does not exist.

We note that the solution to eqs.
(\ref{2.34})-(\ref{2.35}) may not be unique.
The simplest example occurs in the case of one $p$-brane,
when $h_s >0$, $\eps_s = +1$ and $\bar \mu^2 h_s > Q_s^2$.
In this case we have two solutions to (\ref{2.34})-(\ref{2.35})
corresponding to two possible values of $\dot{q^s}(0)$.

{\bf Hypothesis.}
{\em For positive-definite matrix $(h_s A_{s s'})$  and $\eps_s = -1$,
$s \in S$, the solution to (\ref{2.34})-(\ref{2.35}) is
uniquely defined.}

This hypothesis will be a subject of a future investigation.
It implies a "no-hair theorem" for black hole solutions under
consideration.

Thus, we obtained a family of black hole solutions
up to solutions of radial equations (\ref{2.34})
with the boundary conditions (\ref{2.35a}) and  (\ref{2.35}).
In the next sections we consider several exact solutions
to eqs. (\ref{2.34})-(\ref{2.35}).

{\bf Remark 3.} {\em Let $M_i = \R$ and $g^i = -d \bar t \otimes d\bar t$
for some $i \geq 3$. Then the metric (\ref{2.30}) has no a horizon
with respect to the ``second time''  $\bar t$ for $R^{\bar{d}} \to 2\mu$.
Thus, we a led to a ``single-time'' theorem from \cite{Br2}.
Relation (\ref{2.18}) from Proposition 1 coincides with
the ``no-hair'' theorem from \cite{Br2}. }

\section{Polynomial structure of $H_s$ for  Lie algebras}

\subsection{Conjecture on polynomial structure}

Now we deal  with solutions to second order non-linear
differential equations  (\ref{2.34}) that may be rewritten
as follows
\beq{3.1}
 \frac{d}{dz} \left( \frac{F}{H_s}  \frac{d}{dz} H_s \right) = \bar B_s
\prod_{s' \in S}  H_{s'}^{- A_{s s'}},
\eeq
where $H_s(z) > 0$, $F =  1 - 2\mu z$, $\mu > 0$,
$z = R^{-\bar d}$, $\bar B_s =  B_s/ \bar d^2 \neq 0$.
Eqs. (\ref{2.35}) and  (\ref{2.35a}) read
\bear{3.2a}
H_{s}((2\mu)^{-1} -0) = H_{s0} \in (0, + \infty), \\
\label{3.2b}
H_{s}(+0) = 1,
\ear
$s \in S$. (Here we repeat equations (\ref{0.1})-(\ref{0.3}))
for convenience.)

It seems rather difficult to find the solutions to a set
of eqs. (\ref{3.1})-(\ref{3.2b}) for arbitrary
values of parameters $\mu$, $\bar B_s$, $s \in S$ and
quasi-Cartan matrices $A =(A_{s s'})$. But we may
expect a drastically simplification of the problem
under consideration for certain class of parameters and/or
$A$-matrices.

In general we may try to seek solutions of (\ref{3.1})  in a class
of functions analytical in a disc $|z| < L$ and continuous
in semi-interval $0 < z \leq (2\mu)^{-1}$. For $|z| < L$
we get
\beq{3.3}
H_{s}(z) = 1 + \sum_{k = 1}^{\infty} P_s^{(k)} z^k,
\eeq
where $P_s^{(k)}$ are constants, $s \in S$. Substitution
of (\ref{3.3})  into (\ref{3.1}) gives us an infinite
chain of relations on parameters $P_s^{(k)}$  and
$\bar B_s$.  In general case it seems to be impossible
to solve this chain of equations.

Meanwhile there exist solutions to eqs. (\ref{3.1})-(\ref{3.2b})
of polynomial type. The simplest example occurs in orthogonal
case \cite{CT,AIV,Oh,IMJ,BIM}, when
\beq{3.4}
(U^s,U^{s'})= B_{s s'} = 0,
\eeq
for  $s \neq s'$, $s, s' \in S$. In this case
$(A_{s s'}) = {\rm diag}(2,\ldots,2)$ is a Cartan matrix
for semisimple Lie algebra $A_1 \oplus  \ldots  \oplus  A_1$
and
\beq{3.5}
H_{s}(z) = 1 + P_s z,
\eeq
with $P_s \neq 0$, satisfying
\beq{3.5a}
P_s(P_s + 2\mu) = -\bar B_s,
\eeq
$s \in S$.

In \cite{Br,IMJ2,CIM} this solution
was generalized to a block orthogonal
case:
\ber{3.6}
S=S_1 \cup\dots\cup S_k, \qquad  S_i \cap S_j = \emptyset, \quad i \neq j,
\eer
$S_i \ne \emptyset$, i.e. the set $S$ is a union of $k$ non-intersecting
(non-empty) subsets $S_1,\dots,S_k$,
and
\ber{3.7}
(U^s,U^{s'})=0
\eer
for all $s\in S_i$, $s'\in S_j$, $i\ne j$; $i,j=1,\dots,k$.
In this case (\ref{3.5}) is modified as follows
\beq{3.8}
H_{s}(z) = (1 + P_s z)^{b_0^s},
\eeq

where $b_0^s$ are defined in  (\ref{2.20}) and parameters $P_s$
are coinciding inside blocks, i.e. $P_s = P_{s'}$ for $s, s' \in
S_i$, $i =1,\dots,k$. Parameters $P_s \neq 0 $ satisfy the
relations

 $$P_s(P_s + 2\mu) = - \bar B_s/b_0^s,$$

$b_0^s \neq 0$, and parameters $\bar B_s/b_0^s$  are also
coinciding inside blocks, i.e. $\bar B_s/b_0^s = \bar
B_{s'}/b_0^{s'}$ for $s, s' \in S_i$, $i =1,\dots,k$.
In this case
$H_s$ are analytical in $|z| < L$, where $L = {\rm
 min} (| P_ s|^{-1}, s \in S$).

Let $(A_{s s'})$ be  a Cartan matrix  for a  finite-dimensional
semisimple Lie  algebra $\cal G$. In this case all powers in
(\ref{2.20})  are  natural numbers  \cite{GrI}
\beq{3.11}
b_0^s = 2 \sum_{s' \in S} A^{s s'} = n_s \in \N,
\eeq
and  hence, all functions $H_s$ are polynomials, $s \in S$.

Integers $n_s$ coincide with the components  of twice the dual
Weyl vector in the basis of simple coroots (see Sect. 13.7 in
\cite{FS}).

{\bf Conjecture.} {\em Let $(A_{s s'})$ be  a Cartan matrix
for a  semisimple finite-dimensional Lie algebra $\cal G$.
Then  the solution to eqs. (\ref{3.1})-(\ref{3.2b})
(if exists) is a polynomial
\beq{3.12}
H_{s}(z) = 1 + \sum_{k = 1}^{n_s} P_s^{(k)} z^k,
\eeq
where $P_s^{(k)}$ are constants,
$k = 1,\ldots, n_s$, integers $n_s = b_0^s$ are
defined in (\ref{3.11}) and $P_s^{(n_s)} \neq 0$,  $s \in S$}.

In extremal case ($\mu = + 0$) an a analogue of this conjecture
was suggested previously in \cite{LMMP}.

\subsection{Proof of Conjecture  for $A_m$ and $C_{m+1}$ }

First, we prove the {\bf Conjecture} for
simple Lie algebras $A_{m}= sl(m+1)$, $m \geq 1$.
Let us  consider exact solutions to equations of motion
of a Toda-chain corresponding to the Lie algebra
 $A_{m}$ \cite{T,And} ,
\beq{B.1}
\ddot q^s  = - B_s \exp\left( \sum_{s'=1}^{m} A_{s s'} q^{s'}  \right) ,
\eeq
where
\beq{B.1a}
\left(A_{ss'}\right)=
\left( \begin{array}{*{6}{c}}
2&-1&0&\ldots&0&0\\
-1&2&-1&\ldots&0&0\\
0&-1&2&\ldots&0&0\\
\multicolumn{6}{c}{\dotfill}\\
0&0&0&\ldots&2&-1\\
0&0&0&\ldots&-1&2
\end{array}
\right)\quad
\eeq
is the Cartan matrix  of the Lie algebra $A_{m}$ and $B_s > 0$,
$s,s' = 1, \ldots, m$.  Here we put
$S =  \{1, \ldots, m \}$.

The equations of motion   (\ref{B.1}) correspond
to the Lagrangian
\beq{B.2}
 L_T = \frac{1}{2} \sum_{s,s'=1}^{m} A_{ss'} \dot q^s  \dot q^{s'}  -
\sum_{s=1}^{m}  B_s \exp \left( \sum_{s'=1}^{m} A_{ss'} q^{s'}  \right).
\eeq
This Lagrangian may be obtained from the standard one \cite{T}
by separating a coordinate describing the  motion  of the center of mass.

Using the result of A. Anderson \cite{And}
we present the solution to eqs. (\ref{B.1}) in the following form
\beq{B.3}
C_s \exp(-q^s(u)) =
\sum_{r_1< \dots <r_s}^{m+1} v_{r_1}\cdots v_{r_s}
\Delta^2( w_{r_1}, \ldots, w_{r_s}) \exp[(w_{r_1}+\ldots +w_{r_s})u],
\eeq
$s = 1, \ldots, m$, where
\beq{B.4a}
\Delta( w_{r_1}, \ldots, w_{r_s})  =
\prod_{i<j}^{s} \left(w_{r_i}-w_{r_j}\right); \quad
\Delta(w_{r_1}) \equiv 1,
\eeq
denotes the  Vandermonde determinant.
The real constants $v_r$ and $w_r$, $r = 1, \ldots, m + 1$, obey the
relations
\beq{B.5}
\prod_{r=1}^{m+1} v_r= \Delta^{-2}(w_1,\ldots, w_{m+1}), \qquad
\sum_{r=1}^{m+1}w_r=0.
\eeq

In (\ref{B.3})
 \beq{B.6}
 C_s = \prod_{s'=1}^{m} B_{s'}^{-A^{s s'}},
 \eeq
 where
\ber{B.7}
 A^{s s'}= \frac1{m+1}\min(s,s')[m+1 - \max(s,s')],
 \eer
 $s, s' = 1, \ldots, m$,  are components of a matrix inverse to the
 Cartan one, i.e. $(A^{s s'})=(A_{ss'})^{-1}$ (see Sect. 7.5 in
 \cite{FS}).

Here
 \beq{B.8} v_r \neq 0, \qquad w_r \neq w_{r'}, \quad r \neq
  r',
 \eeq
 $r, r' = 1, \ldots, m +1$. We note that the solution with
 $B_s > 0$ may be obtained from the solution  with $B_s =1$ (see
\cite{And}) by a certain shift $q^s \mapsto q^s + \delta^s$.

The energy reads \cite{And}
\beq{B.9}
 E_T= \frac{1}{2} \sum_{s,s'=1}^{m} A_{s s'} \dot q^s  \dot q^{s'} +
\sum_{s=1}^{m}  B_s \exp \left( \sum_{s'=1}^{m} A_{ss'} q^{s'}  \right) =
\frac{1}{2}\sum_{r=1}^{m+1} w^2_r.
\eeq

If  $B_s > 0$, $s \in S$, then  all $w_r, v_r$ are real and, moreover, all
$v_r > 0$, $r =1, \ldots, m+1$.  In a general case $B_s \neq 0$, $s \in
S$, relations  (\ref{B.3})-(\ref{B.6}) also describe real solutions to
eqs. (\ref{B.1}) for suitably chosen complex
parameters $v_r$ and $w_r$. These parameters are either real or belong to
pairs of complex conjugate (non-equal) numbers, i.e., for example, $w_1 =
\bar w_2$, $v_1 = \bar v_2 $. When some of $B_s$ are negative, there are
also some special (degenerate) solutions to eqs. (\ref{B.1}) that are not
described by relations (\ref{B.3})-(\ref{B.6}), but may be obtained from
the latter by certain limits of parameters $w_r$.

For the energy (\ref{1.31}) we get
\beq{B.10}
E_{TL} = \frac{1}{2K} E_T=  \frac{h}{4} \sum_{r=1}^{m+1} w^2_r.
\eeq
Here
\beq{B.11}
K_s = K,    \qquad  h_s= h = K^{-1},
\eeq
$s \in S$.
Thus, in the $A_m$ Toda chain case
eqs.  (\ref{B.3})-(\ref{B.11}) should be substituted into
relations  (\ref{1.19}) and  (\ref{1.30a}).

Now we consider $A_m$-solutions with asymptotics (\ref{2.7}).
In this case all $w_1, \ldots, w_{m+1}$ are real and without
loss of generality $w_1 < \ldots < w_{m+1}$.
For $b_0^s= n_s$  from (\ref{2.20}) we get \cite{GrI}
\beq{B.12}
n_s = b_0^s = s(m-s+1),
\eeq
$s = 1, \ldots, m$, or explicitly
\beq{B.13}
b_0^1 = m, \quad b_0^2 = 2 (m -1), \ldots, b_0^m = m.
\eeq
From (\ref{2.7}), (\ref{2.13}), $\bar{\mu} = \sqrt{C_1}$
and (\ref{B.3}) we get ($w_1 < \ldots < w_{m+1}$)
\bear{B.14}
\bar \mu b_0^1 = \bar \mu m = w_{m+1}, \\ \label{B.15}
\bar \mu b_0^2 = 2 \bar \mu(m-1) =  w_{m} + w_{m+1}, \\ \nn
\ldots \\  \label{B.16}
\bar \mu b_0^m = \bar \mu m =  w_{2}+ \dots + w_{m+1}.
\ear
These relations imply
\beq{B.17}
w_{m+1} = \bar \mu m, \quad  w_{m} = \bar \mu (m -2),
\ldots, w_{1} = - \bar \mu m,
\eeq
or,
\beq{B.18}
w_{j} = (2j - m -2) \bar \mu,
\eeq
$j = 1, \ldots, m + 1$. From (\ref{B.3}) and (\ref{B.17})
we get
\beq{B.19}
 f_{s} = e^{-q^s} = \alpha_s^{(0)} e^{n_s \bar \mu u} +
 \alpha_s^{(1)} e^{(n_s-2) \bar \mu u} +
 \ldots +  \alpha_s^{(n_s)} e^{-n_s \bar \mu u},
\eeq
where $\alpha_s^{(k)}$ are constants, $k = 1, \ldots, n_s$,
$\alpha_s^{(n_s)} \neq 0$. Hence, due to (\ref{2.28}), (\ref{2.28a})
we obtain  the relations
\beq{B.19a}
H_{s} = e^{-q^s- n_s \bar \mu u } =
\alpha_s^{(0)}  + \alpha_s^{(1)} F + \ldots +  \alpha_s^{(n_s)} F^{n_s},
\eeq
equivalent to  (\ref{3.12})
($\alpha_s^{(0)}  + \alpha_s^{(1)} + \ldots +  \alpha_s^{(n_s)} =1)$
with $\alpha_s^{(n_s)} = P_s^{(n_s)}
\neq 0$, $s = 1, \ldots, m$. Thus, the Conjecture is proved for
the Lie algebras ${\cal G} = A_m$, $m \geq 1$.

Now we prove the {\bf Conjecture } for
simple Lie algebras $C_{m+1}= sp(m+1)$, $m \geq 1$.
(Remind that for $m = 1$: $C_2 = B_2 = so(5)$).
The Cartan matrix for the Lie algebra
$C_{m+1}$ ($m \geq 1$) reads
\beq{C.1a}
\left(A_{ss'}\right)=
\left( \begin{array}{*{6}{c}}
2&-2&0&\ldots&0&0\\
-1&2&-1&\ldots&0&0\\
0&-1&2&\ldots&0&0\\
\multicolumn{6}{c}{\dotfill}\\
0&0&0&\ldots&2&-1\\
0&0&0&\ldots&-1&2
\end{array}
\right)\quad
\eeq
$s,s' = 0, \ldots, m$. The set of  equations  (\ref{3.1})
with the Cartan matrix (\ref{C.1a}) and
$s = 0, \ldots, m$, may be embedded into a set
of equations (\ref{B.1}) corresponding to the  Cartan matrix
of the Lie algebra $A_{2m+1}$ (see (\ref{B.1a}))
with  $s = -m, \ldots, 0, \ldots, m$, if the following
identifications : $\bar B_{-k} = \bar B_{k}$
and  $H_{-k} = H_{k}$, $k = 1, \ldots, m$, are adopted.
This proves the Conjecture for $C_{m+1}$,
since it was proved for $A_{2m+1}$.

\section{Some examples}

\subsection{Solution for $A_2$}

Here we consider some examples of solutions related
to the Lie algebra $A_2 = sl(3)$. According to the
results of previous section we  seek the solutions
to eqs. (\ref{3.1})-(\ref{3.2b}) in the following
form (see  (\ref{3.12}); here $n_1 = n_2 =2$):
\beq{4.1}
H_{s} = 1 + P_s z + P_s^{(2)} z^{2},
\eeq
where $P_s= P_s^{(1)}$ and $P_s^{(2)} \neq 0$ are constants,
$s = 1,2$.

The substitution of  (\ref{4.1}) into equations  (\ref{3.1})
and decomposition in powers of $z$ lead us to the relations
\bear{4.2}
 - P_s (P_s + 2 \mu )  + 2 P_s^{(2)} = \bar B_s, \\  \label{4.3}
 - 2 P_s^{(2)} (P_s + 4 \mu ) =  P_{s+1} \bar B_s, \\ \label{4.4}
 - 2 P_s^{(2)} (\mu P_s + P_{s}^{(2)}) =  P_{s+1}^{(2)} \bar B_s,
\ear
corresponding to powers $z^0, z^1, z^2$ respectively, $s = 1,2$.
Here we denote  $s+ 1 = 2, 1$ for $s = 1,2$ respectively.
For $P_1 +P_2 + 4\mu \neq 0$ the solutions of
(\ref{4.2})-(\ref{4.4}) read
\bear{4.5}
 P_s^{(2)} = \frac{ P_s P_{s +1} (P_s + 2 \mu )}{2 (P_1 +P_2 + 4\mu)},
 \\ \label{4.6}
\bar B_s = - \frac{ P_s (P_s + 2 \mu )(P_s + 4 \mu )}{P_1 +P_2 + 4\mu},
\ear
$s = 1,2$. For $P_1 +P_2 + 4\mu = 0$ there exist also a
special solution with
\bear{4.6a}
P_1= P_2 = -2 \mu, \qquad  2 P_{s}^{(2)} = \bar B_s >0,
\qquad \bar B_1 + \bar B_2 = 4 \mu^2.
\ear

Thus, in the $A_2$-case the solution is described by relations
(\ref{2.30})-(\ref{2.33}) with $S = \{s_1,s_2\}$,
intersection rules (\ref{1.40}), or, equivalently,
\bear{1.40a}
d(I_{s_1} \cap I_{s_2})= \frac{d(I_{s_1})d(I_{s_2})}{D-2}-
\chi_{s_1} \chi_{s_2} \lambda_{a_{s_1}}\cdot\lambda_{a_{s_2}}
- \frac12 K,
\\ \label{1.40b}
d(I_{s_i}) - \frac{(d(I_{s_i}))^2}{D-2}+
\lambda_{a_{s_i}}\cdot\lambda_{a_{s_i}} = K,
\ear
where
$K = K_{s_i} \neq 0$,  and functions $H_{s_i} = H_i$
are defined by relations
(\ref{4.1}) and (\ref{4.5})-(\ref{4.6a}) with $z = R^{-\bar d}$,
$i =1,2$.

\subsection{$A_2$-dyon in $D = 11$ supergravity}

Consider the  ``truncated''  bosonic sector of
$D=   11$ supergravity (``truncated''  means without
Chern-Simons term).  The action  (\ref{1.1}) in this
case reads  \cite{CJS}
\ber{4.7}
S_{tr} =   \int_{M} d^{11}z \sqrt{|g|}
\left\{ {R}[g] - \frac{1}{4!}  F^2 \right\}.
\eer
where ${\rm rank} F =   4$. In this particular case,
we consider a dyonic black-hole solutions
with  electric $2$-brane and magnetic  $5$-brane
defined on the manifold
\beq{4.8}
M =    (2\mu, +\infty )  \times
(M_1 = S^{2})  \times (M_2 = \R) \times M_{3} \times M_{4},
\eeq
where ${\dim } M_3 =  2$ and ${\dim } M_4 =  5$.

The solution reads,
\bear{4.9}
g=  H_1^{1/3} H_2^{2/3} \biggl\{ \frac{dR \otimes dR}{1 - 2\mu / R} +
R^2  d \Omega^2_{2} \\ \nn
 -  H_1^{-1} H_2^{-1} \left(1 - \frac{2\mu}{R} \right) dt\otimes dt
+ H_1^{-1} g^3 + H_2^{-1} g^4 \biggr\}, \mm
\label{4.10}
F =  - \frac{Q_1}{R^2} H_1^{-2} H_2  dR \wedge dt \wedge \tau_3+
Q_2 \tau_1 \wedge \tau_3,
\ear
where metrics $g^2$ and  $g^3$ are
Ricci-flat metrics of Euclidean signature,
and $H_s$  are defined as follows
\beq{4.11}
H_{s} = 1 + \frac{P_s}{R}+ \frac{P_s^{(2)}}{R^{2}},
\eeq
where parameters
$P_s$, $\mu > 0$ and $P_s^{(2)}$, $\bar B_s = B_s = - 2 Q_s^2$,
$s =1,2$, satisfy relations  (\ref{4.5}) and (\ref{4.6}).

The  solution describes $A_2$-dyon consisting
of electric  $2$-brane with world sheet isomorphic
to $(M_2 = \R) \times M_{3}$ and magnetic  $5$-brane
with worldsheet isomorphic to $(M_2 = \R) \times M_{4}$.
The ``branes'' are intersecting on the time manifold $M_2 = \R$.
Here  $K_s = (U^s,U^s)=2$, $\eps_s = -1$ for all $s \in S$.
The $A_2$ intersection rule reads (see (\ref{1.40}))
\beq{4.12}
2 \cap 5= 1
\eeq
Here and in what follows $(p_1 \cap p_2= d) \Leftrightarrow
(d(I)=p_1 + 1, d(J)= p_2 + 1, d(I\cap J) = d)$.

The solution (\ref{4.9}), (\ref{4.10})
satisfies not only equations of
motion for the truncated model,
but also  the equations of motion
for  $D =11$ supergravity with the bosonic sector action
\ber{4.13}
S =  S_{tr} +  c \int_{M} A \wedge F \wedge F
\eer
($c = {\rm const}$,  $F = d A$),
since the only modification
related to ``Maxwells'' equations
\ber{4.14}
d*F = {\rm const} \ F \wedge F,
\eer
is trivial due to $F \wedge F = 0$ (since $\tau_i \wedge \tau_i =0$).

This solution in a special case $H_1 = H_2 = H^2$
($P_1 = P_2$, $Q_1^2 =  Q_2^2$) was considered in \cite{CIM}.
The 4-dimensional section of the metric (\ref{4.9})
in this special case coincides
with the Reissner-Nordstr\"om  metric.
For the extremal case, $\mu \to + 0$,
and multi-black-hole generalization
see also  \cite{IMBl}.

\subsection{$A_2$-dyon in Kaluza-Klein model}

Let us  consider $4$-dimensional model
\beq{4.15}
S= \int_{M} d^4z \sqrt{|g|}\biggl\{R[g]- g^{\mu \nu}
\p_\mu \varphi \p_\nu \varphi
-\frac{1}{2!} \exp[2\lambda \varphi]F^2\biggr\}
\eeq
with scalar field  $\varphi$, two-form $F = d A$ and
\beq{4.16}
\lambda = - \sqrt{3/2}.
\eeq
This model originates after Kaluza-Klein reduction
of $5$-dimensional gravity. The 5-dimensional metric
in this case reads
\beq{4.16a}
g^{(5)} = \phi g_{\mu \nu} dx^{\mu} \otimes dx^{\nu}
          + \phi^{-2} (dy + {\cal A}) \otimes (dy + {\cal A}),
\eeq
where
\beq{4.16b}
{\cal A} = \sqrt{2} A =  \sqrt{2} A_{\mu} dx^{\mu},
\qquad \phi = \exp(2 \varphi/\sqrt{6}).
\eeq

We consider the dyonic black-hole solution
carrying  electric  charge $Q_1$
and magnetic  charge $Q_2$, defined on the manifold
\beq{4.17}
M =    (2\mu, +\infty )  \times (M_1 =S^{2})  \times (M_2 = \R).
\eeq
This solution reads
\bear{4.18}
g= \left( H_1 H_2 \right)^{1/2}
\biggl\{ \frac{dR \otimes dR}{1 - 2\mu / R} +   R^2  d \Omega^2_{2}
 -  H_1^{-1} H_2^{-1} \left(1 - \frac{2\mu}{R } \right) dt\otimes dt
\biggr\}, \\ \label{4.19}
\exp(\varphi) = H_1^{\lambda/2} H_2^{- \lambda/2},
\\ \label{4.20}
F = dA = -  \frac{Q_1}{R^2} H_1^{-2} H_2  dR \wedge dt + Q_2 \tau_1,
\ear
where  functions $H_{s}$  are defined by relations (\ref{4.1}),
(\ref{4.5}) and (\ref{4.6}) with
$\bar B_s = - 2 Q_s^2$, $z = R^{- 1}$, $s =1, 2$;
where $\tau_1$ is volume form on $S^2$.

For 5-metric we obtain from (\ref{4.16a})-(\ref{4.19})
\bear{4.16c}
g^{(5)} = H_2 \biggl\{ \frac{dR \otimes dR}{1 - 2\mu / R}
+   R^2  d \Omega^2_{2}   -  H_1^{-1} H_2^{-1}
\left(1 - \frac{2\mu}{R } \right) dt\otimes dt \biggr\}
\\ \nn
 + H_1 H_2^{-1}  (dy + {\cal A}) \otimes (dy + {\cal A}),
\ear
$d {\cal A} = \sqrt{2} F$.

For $Q_2 \to 0$ we get the black hole version of
Dobiash-Maison solution from \cite{DoMa} and
for $Q_1 \to 0$ we are led to the black hole version of
Gross-Perry-Sorkin monopole solution from \cite{GrP,Sor},
see \cite{CGMS}.
The solution coincides  with
Gibbons-Wiltshire dyon solution  \cite{GW}. Our notations are related to
those from ref. \cite{GW}, as following :  $H_1 R^2 = B$, $H_2 R^2 = A$,
$R^2 - 2 \mu R = \Delta$, $Q_1 = \sqrt{2} q$, $Q_2 = - \sqrt{2} p$, $R
-\mu = r- m$, $\mu^2 = m^2 + d^2 - p^2 -q^2$, $(P_2 -P_1)/2(P_2 +1) = d/(d
-\sqrt{3} m)$).  (For general spherically symmetric configurations see
also ref. \cite{Lee}.)

We note that, quite recently, in \cite{CGS} the KK dyon solution
\cite{GW} was used for constructing the dyon solution in $D=11$
supergravity (\ref{4.9})-(\ref{4.10}) for flat $g^3$ and $g^4$ and its
rotating version.

\section{Conclusions}

Thus here we obtained a family
of black hole (BH) solutions  with intersecting $p$-branes
with nearly arbitrary intersection rules,
see relations (\ref{1.17a}), (\ref{1.18b}) and
Restriction 1. (Restriction 2 is satisfied, since all
$p$-branes have a common time manifold.)
These BH solutions are given by relations
(\ref{2.30})-(\ref{2.35}).
The metric of solutions  contains  $n -1$ Ricci-flat ``internal''
space metrics. The solutions are defined up to  a set of
("moduli") functions $H_s$  obeying a set of  equations
(\ref{2.34}) with  boundary conditions (\ref{2.35a})-(\ref{2.35})
(or, equivalently, eqs. (\ref{0.1})-(\ref{0.3})).

These solutions are new and generalize a lot of
special classes of BH solutions considered earlier in the literature.
It is not necessary in future investigations to consider
special models and setups,  find spherically symmetric solutions and
single out BH ones. All this program is fulfilled in this paper
(with the use of results of ref. \cite{IKM}). What we only  need is to
find explicit relations for moduli functions $H_s$, when  matrix $A$ is
fixed, i.e.  to solve equations (\ref{0.1}) with  boundary conditions
(\ref{0.2})-(\ref{0.3}) imposed.
The problem (\ref{0.1})-(\ref{0.3})  seems to be
rather difficult and  may be of  interest from the pure mathematical
point of view, regardless to possible physical applications.

Here we suggested a conjecture on polynomial structure of  $H_s$
for intersections related to semisimple Lie algebras and
proved it for $A_m$ and $C_{m+1}$ algebras, $m \geq 1$.
This result may be interesting, since
any appearance of polynomials in mathematical physics,
especially related to Lie algebras, is always a rather
attractive  for mathematicians (and physicists, as well).

Here we also obtained explicit relations for the solutions in the
$A_2$-case and considered two examples of $A_2$-dyon solutions:
one in $D = 11$  supergravity (with $M2$- and $M5$-branes intersecting at
a point ) and another in $5$-dimensional Kaluza-Klein theory
(Gibbons-Wiltshire solution).  Explicit relations  for $H_s$ corresponding
to other examples of Lie algebras (e.g.  $A_3$, $B_2$ etc) will be
considered in future publications.

\begin{center}
{\bf Acknowledgments}
\end{center}

This work was supported in part by the Russian Ministry for
Science and Technology, Russian Foundation for Basic Research,
and project SEE.


\small
\begin{thebibliography}{99}

\bibitem{M-th1}
E. Witten, {\it Nucl. Phys.} {\bf B 443}, 85 (1995); hep-th/9503124; \\
P. Townsend, {\it Phys. Lett. } {\bf B 350}, 184 (1995); hep-th/9612121; \\
C. Hull and P. Townsend, {\it Nucl. Phys.} {\bf B 438}, 109 (1995);
hep-th/9610167; \\
P. Horava  and E. Witten, {\it Nucl. Phys.} {\bf B 460}, 506 (1996);
hep-th/9510209.

\bibitem{M-th2}
J.M. Schwarz,  Lectures on Superstring and M-theory Dualities,
hep-th/9607201; \\
M.J. Duff,  M-theory (the Theory Formerly Known as Strings),
hep-th/9608117.


\bibitem{GSW}
M.B. Green, J.H. Schwarz and E. Witten, Superstring Theory,
vol. 1, 2, Cambridge, 1987.

\bibitem{St}
K.S. Stelle, Lectures on Supergravity p-branes, hep-th/9701088.

\bibitem{DKL}
M.J. Duff, R.R. Khuri and J.X. Lu,
{\it Phys. Rep.} {\bf 259},  213 (1995).

\bibitem{DGHR}
A. Dabholkar, G. Gibbons, J.A. Harvey and F. Ruiz Ruiz, {\it Nucl.
Phys.} {\bf B 340}, 33 (1990).

\bibitem{HS}
G.T. Horowitz  and A. Strominger,
{\it Nucl. Phys.} {\bf B 360}, 197 (1990).

\bibitem{DS}
M.J. Duff and K.S. Stelle, {\it Phys. Lett.} {\bf B 253},  113 (1991).

\bibitem{Guv}
R. G\"{u}ven, Phys. Lett. {\bf B 276}, 49 (1992);
{\it Phys. Lett.} {\bf B 212}, 277 (1988).

\bibitem{Str}
A. Strominger, {\it Phys. Lett. } {\bf B 383}, 44 (1996);
hep-th/9512059.

\bibitem{To}
P.K. Townsend, {\it Phys. Lett. } {\bf B 373}, 68 (1996);
hep-th/9512062.

\bibitem{PT}
G. Papadopoulos and P.K. Townsend,
{\it  Phys. Lett.} {\bf B 380}, 273 (1996); hep-th/9603087.

\bibitem{Ts}
A.A. Tseytlin,
{\it Nucl. Phys.} {\bf B 475}, 149 (1996); hep-th/9604035.

\bibitem{GKT}
J.P. Gauntlett,  D.A. Kastor, and J. Traschen,
{\it Nucl. Phys.} {\bf B 478}, 544 (1996); hep-th/9604179.

\bibitem{CT}
M. Cvetic and A.A. Tseytlin,  Nucl. Phys. B 478, 181 (1996).

\bibitem{Ts1}
A.A. Tseytlin, {\it Nucl. Phys.} {\bf B 487}, 141 (1997);
hep-th/9609212.

\bibitem{LP}
H. L\"u, C.N. Pope, SL(N+1,R) Toda Solitons in Supergravities,
{\it Int. J. Mod. Phys.}  {\bf A 12}, 2061 (1997); hep-th/9607027.

\bibitem{LPX}
H. L\"u, C.N. Pope, and K.W. Xu, Liouville and Toda Solitons in
M-theory,
{\it  Mod. Phys. Lett.}  {\bf A 11}, 1785 (1996); hep-th/9604058.

\bibitem{LMPX}
H. L\"u, S. Mukherji, C.N. Pope and K.-W. Xu,
Cosmological Solutions in String Theories,
{\it  Phys. Rev. }  {\bf D 55}, 7926 (1997); hep-th/9610107.

\bibitem{V}
A. Volovich,  {\it  Nucl. Phys. }  {\bf B 487} (11), 141 (1997);
hep-th/9608095.

\bibitem{AV}
I.Ya. Aref'eva and A.I. Volovich,
{\it Class. Quantum Grav.} {\bf B 14}, 29901 (1997);
hep-th/9611026.

\bibitem{IM0}
V.D. Ivashchuk and V.N. Melnikov,
Intersecting p-brane Solutions in Multidimensional
Gravity and M-theory, hep-th/9612089;
{\it Grav. and Cosmol.} {\bf 2}, No 4, 204 (1996).

\bibitem{IM}
V.D. Ivashchuk and V.N. Melnikov,
{\it Phys. Lett. } {\bf B 403}, 23 (1997).

\bibitem{BREJS}
E. Bergshoeff, M. de Roo, E. Eyras, B. Janssen and
J.P. van der Schaar, {\it Class. Quantum Grav.} {\bf 14} , 2757 (1997);
hep-th/9612095.

\bibitem{AR}
I.Ya. Aref'eva and O.A. Rytchkov,
Incidence Matrix Description of Intersecting p-brane
Solutions, {\it Preprint} SMI-25-96, hep-th/9612236.

\bibitem{AEH}
R. Argurio, F. Englert and L. Hourant,
{\it Phys. Lett. } {B 398}, 2991 (1997); hep-th/9701042.

\bibitem{AIR}
I.Ya. Aref'eva, M.G. Ivanov and O.A. Rytchkov,
Properties of Intersecting p-branes in Various Dimensions,
{\it Preprint} SMI-05-97, hep-th/9702077.

\bibitem{AIV}
I.Ya. Aref'eva, M.G. Ivanov and I.V. Volovich,
Non-Extremal Intersecting p-Branes in Various Dimensions, hep-th/9702079;
{\it Phys. Lett. } {\bf B 406}, 44 (1997).

\bibitem{Oh}
 N. Ohta, Intersection rules for non-extreme
 p-branes, hep-th/9702164; {\it Phys. Lett. } {\bf B 403}, 218-224
(1997).

\bibitem{IMC}
V.D. Ivashchuk and V.N. Melnikov,
Sigma-model for the Generalized  Composite p-branes,
hep-th/9705036; {\it Class. Quantum Grav.} {\bf 14}, 3001 (1997);
Corrigenda {\it ibid.} {\bf 15 } (12), 3941 (1998).

\bibitem{IMR}
V.D. Ivashchuk, V.N. Melnikov and M. Rainer,
Multidimensional $\sigma$-models with Composite Electric $p$-branes,
gr-qc/9705005; {\it Grav. and Cosmol. } {\bf 4}, No 1 (13), (1998).

\bibitem{BGIM}
K.A. Bronnikov, M.A. Grebeniuk, V.D. Ivashchuk and V.N. Melnikov,
Integrable Multidimensional Cosmology for
Intersecting $p$-branes,
{\it Grav. and Cosmol. } {\bf  3}, No 2(10), 105 (1997).

\bibitem{LMMP}
H. L\"u, J. Maharana, S. Mukherji  and C.N. Pope,
Cosmological Solutions, p-branes and the Wheeler De Witt
Equation,  {\it  Phys. Rev. } {\bf D 57}, 2219 (1998);
hep-th/9707182.

\bibitem{GrIM}
M.A. Grebeniuk, V.D. Ivashchuk and V.N. Melnikov,
Integrable Multidimensional Quantum Cosmology  for Intersecting p-Branes,
{\it Grav. and Cosmol.\/} {\bf 3}, No 3 (11), 243 (1997), gr-qc/9708031.

\bibitem{BKR}
K.A. Bronnikov, U. Kasper and M. Rainer,
Interesecting electric and magnetic $p$-branes:
spherically-symmetric solutions, gr-qc/9708058;
{\it GRG}, {\bf 31}, No 11, 1681 (1999).

\bibitem{IMJ}
V.D. Ivashchuk  and  V.N. Melnikov, Multidimensional Classical
and Quantum Cosmology with Intersecting $p$-branes,
{\it J. Math. Phys.}, {\bf 39}, 2866 (1998); hep-th/9708157,

\bibitem{Y}
D. Youm, {\it Phys. Rept.}, {\bf 316} (1999) 1-232;  hep-th/9710046.
(this review on black holes in string theories is rather complete
although it does not contain some important citations,
e.g. \cite{AR}, \cite{AIV}, \cite{Br} etc.)


\bibitem{BIM}
K.A. Bronnikov, V.D. Ivashchuk and V.N. Melnikov,
The Reissner-Nordstr\"om Problem for
Intersecting Electric and Magnetic $p$-branes, gr-qc/9710054;
{\it Grav. and Cosmol.}, {\bf 3}, No 3(11), 203 (1997).

\bibitem{Br}
K.A. Bronnikov, Block-orthogonal Brane systems, Black
Holes and Wormholes, hep-th/9710207;
{\it Grav. and Cosmol.} {\bf 4}, No 1 (13),  49 (1998).

\bibitem{GR}
D.V. Gal'tsov and O.A. Rytchkov, Generating Branes via
Sigma models, hep-th/9801180.

\bibitem{IMBl}
V.D. Ivashchuk and V.N. Melnikov,
Madjumdar-Papapetrou Type Solutions in Sigma-model
and Intersecting p-branes,
{\it Class. Quantum Grav.} {\bf 16}, 849 (1999);
hep-th/9802121.

\bibitem{IKM}
V.D.Ivashchuk, S.-W.Kim and V.N.Melnikov, Hyperbolic Kac-Moody
Algebra from Intersecting $p$-branes,
{\it J. Math. Phys.} {\bf 40}, 4072 (1999);
hep-th/9803006.

\bibitem{GrI}
M.A. Grebeniuk and V.D. Ivashchuk,
Sigma-model Solutions and Intersecting p-branes
Related to Lie Algebras,
{\it Phys. Lett. } {\bf B 442}, 125 (1998);
hep-th/9805113.

\bibitem{Br2}
K.A. Bronnikov,
Gravitating Brane Systems: Some General Theorems,
gr-qc/9806102; {\it J. Math. Phys.} {\bf 40}, 924 (1999).

\bibitem{GM1}
V.R. Gavrilov and V.N. Melnikov,
Toda Chains with Type $A_m$  Lie Algebra for Multidimensional Classical
Cosmology with Intersecting $p$-branes, In :  Proceedings of the
International seminar "Curent topics in mathematical cosmology", (Potsdam,
Germany , 30 March - 4 April 1998), Eds. M. Rainer and H.-J. Schmidt,
World Scientific, 1998,  p. 310; hep-th/9807004.

\bibitem{IMJ2}
V.D. Ivashchuk and V.N. Melnikov,
Multidimensional Cosmological and Spherically Symmetric Solutions
with Intersecting $p$-branes, gr-qc/9901001; \\
Cosmological and Spherically Symmetric Solutions
with Intersecting $p$-branes, {\it  J. Math. Phys.}
{\bf 40}, No 10 (1999), 6558.

\bibitem{CIM}
S. Cotsakis, V.D. Ivashchuk and V.N. Melnikov,
P-branes Black Holes and Post-Newtonian Approximation,
{\it Grav. and Cosmol.\/} {\bf 5}, No 1 (17), 52 (1999);
gr-qc/9902148.

\bibitem{GM2}
V.R. Gavrilov and V.N. Melnikov, Toda Chains  Associated with Lie
Algebras  $A_m$ in Multidimensional Gravitation and Cosmology with
Intersecting $p$-branes, {\it Theor. Math. Phys.} {\bf 123}, No 3,
374-394 (2000) (in Russian).

\bibitem{IK}
V.D. Ivashchuk and S.-W. Kim, Solutions with intersecting p-branes
related to Toda chains, {\it J. Math. Phys.} {\bf 41} (1), 444-460
(2000); hep-th/9907019.

\bibitem{FS}
J. Fuchs and C. Schweigert, Symmetries, Lie algebras and
Representations. A graduate course for physicists
(Cambridge University Press, Cambridge, 1997).

\bibitem{CJS}
E. Cremmer, B. Julia, J. Scherk.
{\it Phys. Lett. } {\bf B 76}, 409 (1978).

\bibitem{MP}
S.D. Majumdar, {\it Phys. Rev. } {\bf 72}, 930 (1947);  \\
A. Papapetrou,  {\it Proc. R. Irish Acad. } {\bf A51}, 191 (1947).

\bibitem{IMZ}
V.D. Ivashchuk, V.N. Melnikov and A.I. Zhuk,
{\it Nuovo Cimento } {\bf B 104}, 575  (1989).

\bibitem{T}
M. Toda, {\it Progr. Theor. Phys.} {\bf 45}, 174 (1970).

\bibitem{And}
A. Anderson, {\it J. Math. Phys.} {\bf 37}, 1349 (1996);
hep-th/9507092.

\bibitem{DoMa}
P. Dobiash and D. Maison,
{\it Gen. Rel. Grav.} {\bf 14}, 231 (1982).

\bibitem{GrP}
D.J. Gross and M.J. Perry,
{\it Nucl. Phys. } {\bf B 226}, 29  (1983).

\bibitem{Sor}
R.D. Sorkin, {\it Phys. Rev. Lett.} {\bf 51}, 87 (1983).

\bibitem{Lee}
S.-C. Lee, {\it Phys. Lett.} {\bf 149}, 98 (1984).

\bibitem{GW}
G. Gibbons and D. Wiltshire, {\it Ann. Phys.} { \bf 167}, 201 (1986);
Erratum: {\it ibid} { \bf 176}, 393 (1987).

\bibitem{CGMS}
C.-M. Chen, D. V. Gal'tsov, K. Maeda and S. Sharakin,
{\it Phys. Lett.} {\bf B 453}, 7 (1999).

\bibitem{CGS}
C.-M. Chen, D. V. Gal'tsov and S. Sharakin, Einstein Gravity ---
Supergravity Correspondence, hep-th/9912127.





\end{thebibliography}
\end{document}